\documentclass[12pt]{amsart}
\usepackage{amsfonts,amssymb,amsmath,amscd}
\usepackage{epsf}

\DeclareFontFamily{U}{rsf}{}
\DeclareFontShape{U}{rsf}{m}{n}{
  <5> <6> rsfs5 <7> <8> <9> rsfs7 <10-> rsfs10}{}
\DeclareMathAlphabet\Scr{U}{rsf}{m}{n}

\newcommand{\al}{{\alpha}}

\renewcommand{\d}{{\rm d}}

\newcommand{\be}{{\beta}}

\renewcommand{\l}{\left}
\renewcommand{\r}{\right}
\def\lr{{\leftrightarrow}}
\newcommand{\LG}{Landau-Ginzburg }
\def\part{\partial}

\newcommand{\Tr}{{\rm Tr}}
\newcommand{\str}{{\rm STr}}

\newcommand{\C}{{\mathsf C}}
\newcommand{\CC}{{\mathbb C}}

\newcommand{\ZZ}{{\mathbb Z}}

\newcommand{\Hom}{{\rm Hom}}

\newcommand{\End}{{\rm End}}

\newcommand{\Mat}{{\rm Mat}}
\newcommand{\Cl}{{\rm Cl}}

\newcommand{\cO}{{\mathcal O}}
\newcommand{\cD}{{\mathcal D}}
\newcommand{\bE}{{\bar E}}

\newcommand{\N}{{\mathcal N}}

\newcommand{\D}{{\mathcal D}}

\renewcommand{\O}{{\mathcal O}}

\newcommand{\cH}{{\Scr{H}}}

\newcommand{\no}{\nonumber}
\newcommand{\op}{\oplus}
\newcommand{\ot}{\otimes}

\def\bea{\begin{eqnarray}}
\def\eea{\end{eqnarray}}
\def\bmat{\begin{pmatrix}}
\def\emat{\end{pmatrix}}

\def\su{{\rm SU}}
\def\tj{\tilde{j}}
\def\tn{\tilde{n}}
\def\tB{\tilde{B}}
\def\ts{\tilde{s}}
\def\drangle{\rangle\!\rangle}
\newcommand{\im}{{\rm Im\,}}

\title[D-branes in Topological Minimal Models]{D-branes in Topological Minimal Models: the Landau-Ginzburg
Approach}
\author[A. Kapustin]{Anton Kapustin \hspace{3mm} }
\address{California Institute of Technology\\
Department of Physics\\
Pasadena, CA 91125}
\email{kapustin@theory.caltech.edu}
\author[Y. Li]{ \hspace{3mm} Yi Li}
\address{California Institute of Technology\\
Department of Physics\\
Pasadena, CA 91125}
\email{yili@theory.caltech.edu}

\begin{document}

\begin{abstract}
We study D-branes in topologically twisted $\N=2$ minimal models using the
Landau-Ginzburg realization. In the cases of $A$ and $D$-type minimal models
we provide what we believe is an exhaustive list of topological branes and compute the
corresponding boundary OPE algebras as well as all disk correlators. We also construct examples of topological
branes in $E$-type minimal models.
We compare our results with the boundary state formalism, where possible, and find 
agreement.

\end{abstract}

\maketitle

\vspace{-5in}

\parbox{\linewidth}
{\small\hfill \shortstack{CALT-68-2412}} 

\vspace{5in}

\section{Introduction}\label{sec:intro}

Topological Landau-Ginzburg models are a rich, and at the same time simple, class of topological field theories (TFTs).
When these models are considered on a world-sheet
without boundaries, all topological correlators are given by a simple closed formula~\cite{Vafa}.
In Refs.~\cite{usone,ustwo} we have generalized these results to world-sheets with boundaries which lie
on arbitrary D-branes of type B. In this paper we work out a particular example: D-branes in
topologically twisted $\N=2$ minimal models. $\N=2$ minimal models are rational $\N=2$ superconformal
field theories. Modular-invariant combinations of characters of $\N=2$ super-Virasoro algebra have ADE
classification, so one can talk about minimal models of type $A$, $D$, or $E$. These theories have Landau-Ginzburg
realizations, the superpotentials being
\begin{align*}
W_{A_m}\;&=\;x^{m+1}  &m\ge1 ,\\
W_{D_m}\;&=\;x^{m-1}+xy^2  &m\ge4,\\
W_{E_6}\;&=\;x^3+y^4,\\
W_{E_7}\;&=\;x^3+xy^3,\\
W_{E_8}\;&=\;x^3+y^5.
\end{align*}
$\N=2$ minimal models admit a B-twist, so one can study associated TFTs and their D-branes. This is what we will do in this paper. Prior
work in this direction includes Refs.~\cite{HIV,Hlinear,Hbook,BH}.\footnote{While
this paper was in preparation, there appeared Ref.~\cite{BHLS}, which also discusses D-branes in $A$-type minimal models using the
approach of Ref.~\cite{usone}.}

One can also study D-branes in the untwisted minimal models using methods of boundary conformal field theory.
The simplest D-branes are those which preserve
the full chiral algebra; these are so-called Cardy branes. In fact, since $\N=2$ super-Virasoro algebra
admits a non-trivial automorphism (the mirror involution), there are two kinds of Cardy branes, which are
exchanged by the mirror involution. They are known as A-branes and B-branes, because they are similar
to A and B-branes on Calabi-Yau manifolds. Cardy B-branes are precisely D-branes compatible
with the topological B-twist, so one can study D-branes in topologically twisted minimal models using the formalism of boundary 
conformal field theory (the boundary state formalism). For minimal models of type $A$, such analysis has been
performed in Ref.~\cite{MMS}, and we will compare our results with those of Ref.~\cite{MMS} below. It should
be stressed that our methods enable one to compute all possible topological correlators, including
those on Riemann surfaces with arbitrary number of handles and holes. This is very hard to do in the
boundary state formalism, where so far only certain disk correlators have been computed. On the other hand,
the boundary state formalism allows one to access non-topological correlators, which are beyond the
reach of our methods. 

To set the stage, let us summarize the main results of Ref.~\cite{Vafa} and Refs.~\cite{usone,ustwo}. Related
results have also appeared in Refs.~\cite{Hlinear,Hbook}.
First, let us recall how to compute bulk topological correlators in a LG model with superpotential $W$. 
To describe them, it is sufficient to 
specify the ring of bulk observables and the one-point function on a genus-$g$ Riemann surface for all $g$.
The bulk ring is given by
$$
\cO/\partial W,
$$
where $\cO$ is the ring of holomorphic functions on the target space $X$.
This is known as the Jacobi ring of $W$. In what follows we will assume that $X=\CC^n$. We also assume that
$W$ is polynomial and replace
the ring of holomorphic functions with the ring of polynomial functions $\CC[z_1,\ldots,z_n]$. Finally, we
will assume that all critical points of $W$ are isolated. In this case the Jacobi ring is finite-dimensional.

Let $\alpha$ be a polynomial representing an element of the Jacobi ring. The corresponding one-point function on a 
genus-$g$ Riemann surface is given by
$$
\langle\alpha\rangle_g=\;\frac1{(2\pi i)^n} \oint \frac{\alpha H^g}{\part_1W\part_2W\ldots\part_nW},
$$
where $H$ is the Hessian of $W$.
In this formula the integrand is regarded as a meromorphic $n$-form (i.e. a factor $dz_1\wedge\ldots\wedge dz_n$ is implied),
and the integral is performed over an $n$-dimensional real submanifold defined by the equations $|\partial_i W|=\epsilon_i,$ where all 
$\epsilon_i$ are small~\cite{Vafa}.
For sufficiently small $\epsilon_i$ this submanifold is a union of several Lagrangian tori each of which encloses a single critical point of $W$.
Alternatively, we may rewrite this expression as an integral of a certain form of type $(n,n-1)$ over a large $2n-1$-dimensional
sphere in $\CC^n$~\cite{GH}. The possibility of such a rewriting expresses the fact that the integral is a kind of
multi-dimensional residue, and can be evaluated using only the behavior of the integrand at infinity.
The above formula is derived by evaluating the path-integral in the zero-mode approximation, which can be argued
to be adequate in the topological sector.

Second, let us describe topological D-branes and the corresponding correlators. Topological D-branes
are localized at the critical points of $W$, and D-branes sitting at different critical points do not ``talk'' to
each other. Thus we may focus on any one critical point. Without loss of generality, we may assume that the critical
point is at $z=0$, and $W(0)=0$. Then D-branes sitting at $z=0$ correspond to ways
of factorizing of $W$ into a product of two matrix polynomials:
$$
d_0 d_1=d_1 d_0=W\cdot id, \quad d_0,d_1\in \Mat(r,\CC[z_1,\ldots,z_n]).
$$
This was proposed by M.~Kontsevich (unpublished). A physical derivation of Kontsevich's proposal has been given
in Refs.~\cite{usone, ustwo}.

Given such matrix factorization, we can describe boundary operators as follows. Consider a $2r\times 2r$ matrix
polynomial 
$$
Q=\begin{pmatrix} 0 & d_1 \\ d_0 & 0\end{pmatrix}.
$$
It can be regarded as a linear operator on the vector space $M$ of $2r$-dimensional vectors with polynomial components.
$M$ has a natural $\ZZ_2$ grading: the first $r$ components are declared even, while the last $r$ are declared odd.
The operator $Q$ is odd, in the sense that it maps the even part of $M$ to its odd part, and vice versa. $Q$ satisfies
$Q^2=W\cdot id.$ In mathematical terminology, $M$ is a free module of rank $2r$ over the algebra of polynomial functions
$\O$. In Ref.~\cite{ustwo}, the pair $(\cO,W)$ is called a CDG algebra, while the pair $(M,Q)$ as above is
called a free CDG module over the CDG algebra. The origin of this terminology is explained in Ref.~\cite{ustwo}.

Now consider the space $P$ of all polynomial linear operators on $M$ which commute with multiplication
by polynomial functions. This is simply the space of $2r\times 2r$
matrices with polynomial entries, and it has an obvious grading. Define a linear operator on $P$ using the ``adjoint'' action
of $Q$:
$$
\cD: a\mapsto [Q,a]=Qa- (-1)^{|a|} a Q, \quad a\in P.
$$
Here $|a|=0$ or $1$ depending on whether $a$ is even or odd. It is easy to see that $\cD$ satisfies
$\cD^2=0$. It was shown in Ref.~\cite{ustwo} that the cohomology of $\cD$ coincides with the space of
physical boundary operators. This space is graded, and it also has the structure of a ring (because one can multiply
matrices). This is nothing but the ring of boundary operators. Thus the boundary ring is completely determined by the
matrix $Q$. One can show that although $P$ is infinite-dimensional, the cohomology of $\cD$ is finite-dimensional,
as expected on physical grounds. In mathematical terms, $\cD$ cohomology computes the endomorphism algebra of a CDG module
(in the derived category of CDG modules). Therefore we will use the expressions ``boundary OPE algebra'' and 
``endomorphism algebra'' interchangeably.

Given a pair of CDG modules, one can also compute the space of boundary-changing operators for the corresponding pair of branes~\cite{ustwo}.
This space can be identified with the space of morphisms in the derived category of CDG modules. Multiplication of
boundary-changing operators corresponds to the composition of morphisms. 

Finally, one can write down a closed formula for arbitrary bulk-boundary topological correlators. Consider a Riemann surface with
$g$ handles and $h$ holes. Suppose the boundary of the $i^{\rm th}$ hole is mapped to a D-brane associated
with a matrix polynomial $Q_i$. Clearly, it is sufficient to consider the case when there is one bulk operator
insertion $\alpha$ and $h$ boundary operator insertions $\phi_i$, one on each boundary circle. Here $\alpha$ is an
element of the Jacobi ring, and $\phi_i$ is a class in the cohomology of $\cD_i$. We have shown~\cite{ustwo} that
the corresponding correlator is given by
$$
\frac1{(n!)^h (2\pi i)^n} \oint \frac{\alpha H^g}{\part_1W\part_2W\ldots\part_nW} \cdot \prod_{i=1}^h
\str\big[(\partial Q_i)^{\wedge n}\phi_i\big].
$$
Here we identify top forms like $(\partial Q_i)^{\wedge n}$ with functions, and an overall factor $dz_1\wedge\ldots\wedge dz_n$ is implied,
so the integrand is a meromorphic $n$-form. The integration is over a union of Lagrangian tori enclosing all the
critical points of $W$, as before. This integral can be regarded as a generalized residue, and can also be rewritten
as an integral of a $2n-1$-form over a large $2n-1$-dimensional sphere in $\CC^n$~\cite{GH}. Some special cases of this formula
have been derived in Refs.~\cite{Hlinear,Hbook}.

\section{D-branes in $\N=2$ minimal models}
\label{sec:minimal}

In this section we classify topological B-branes in $\N=2$ minimal models using the methods described above. 
Previous works which use the \LG viewpoint to study B-branes in minimal models include Refs.~\cite{HIV,Hlinear,Hbook, BHLS}. 
In these papers only $A$-type minimal models have been discussed.
Our computation agrees with the previously obtained results and clarifies some additional subtleties. 
We also provide what we think is a complete list of irreducible B-branes in the case of $D$-type superpotential. 
We conjecture that any other B-brane for the $D$-type
superpotential is a direct sum of the ones we have constructed. Finally we give examples of B-branes in 
$E$-type minimal models.

\subsection{$A$-type minimal models}
\label{subsec:A}

As mentioned above, $\N=2$ minimal models of type $A$ are believed to be the infrared fixed points of \LG models with 
superpotential $W_{A_m}=z^{m+1}$. Before going into detailed calculations, we pause to comment on an important subtlety. Naively, adding a
massive chiral field has no effect in the infrared. This is obviously true in the closed string sector (the Jacobi ring is unaffected by
the addition of squares). In the open string sector, however, 
there is a nontrivial effect. As we shall see, adding a single massive chiral field  leads to a different D-brane spectrum. 
What is the interpretation of this in CFT terms? 

Any $\N=1$ SCFT has a $\ZZ_2$ symmetry which acts non-trivially only on the left-moving Ramond-sector states. In the context of
string theory, this symmetry is usually called $(-1)^{F_L}$.
Orbifolding by this $\ZZ_2$ gives a new $\N=1$ SCFT which has a different spectrum of D-branes. If the original theory has $\N=2$
superconformal symmetry, so will the orbifolded one. We claim that adding an extra massive
chiral field to the LG superpotential has exactly the same effect on D-branes as orbifolding by this $\ZZ_2$ symmetry. To test this claim, note
that the orbifolded model also has a $\ZZ_2$ symmetry, which acts by reversing the sign of all twisted sector states, which are
all in the Ramond sector. Orbifolding by this
$\ZZ_2$ gives the original (unorbifolded) CFT. On the LG side, the second orbifolding corresponds to adding yet another free massive 
chiral superfield,
therefore we expect that adding two squares to $W$ has no effect on D-branes. This is indeed true, for arbitrary $W$~\cite{Orlov}.
Below, we will verify our claim in the case of $A$-type minimal models.

With this ambiguity in mind, we will classify in the following the B-branes for both $W=z^n$ and $W=z^n+y^2$. We will see in the next section 
that they match up precisely with the Cardy branes in the minimal model and its $\ZZ_2$ orbifold. The D-brane spectrum for $W=z^n$ has already 
appeared in the literature \cite{Orlov,BHLS}, so we start by summarizing this case first. As explained in Ref.~\cite{usone,ustwo} and reviewed in the 
Introduction, 
B-branes are classified by 
CDG modules $(E,Q)$ over the $\ZZ_2$-graded CDG algebra $(\O,W)$. In our case $\O$ is simply the algebra of polynomials in a single variable $z$.
We have $k+1$ obvious solutions to the equation $Q^2=W$:
\begin{equation}
E_k \;=\; \l\{\O\oplus\O, \;Q = \begin{pmatrix}0&z^k\\z^{n-k}&0\end{pmatrix}\r\},\label{eq:Ek_A}
\end{equation}
where $k=0,\ldots,n.$ Recall that the BRST operator acts on $P_k\simeq \Mat(2,\CC[x])$ as
$$\D:\;\phi\mapsto [Q,\phi]$$
and the space of endomorphisms (topological open strings) on the brane $E_k$ is given by the $\D$-cohomology. The cases $k=0$ and $k=n$ actually 
turn out 
trivial: the $\D$ cohomology vanishes, as one can easily check.\footnote{Alternatively, this can be seen from the path-integral derivation of 
Kontsevich's proposal in Ref.~\cite{ustwo}, 
since the path integral localizes at $Q=0$. For $k=0$ or $k=n$ this equation has no solutions, and the path integral identically vanishes.} 
In mathematical terms, 
this means that $E_0$ and $E_n$ are zero objects in the category of B-branes. Interesting branes correspond to the range $0<k<n$. 
In fact one can further restrict
the range to $0<k\le n/2$ for the following reason. To any brane one can associate its anti-brane by exchanging $d_0$ and $d_1$ and flipping the
grading on $E$. We will call this operation parity-reversal and will denote the parity-reversal of $E$ by $\bE$. Clearly, 
the space of topological 
strings between a brane $E$ and some other brane $E'$ is the same as the space of topological strings between $\bE$ and $E'$, except for the
reversal of grading. In our case parity-reversal amounts to $k\mapsto n-k$. Thus it is sufficient to deal with $k$ in the range $0<k\le n/2.$
A case of special importance is when $n$ is even and $k=n/2$. This brane is its own anti-brane.

It is straightforward to check that for $k\leq n/2$ the space of endomorphisms of $E_k$ 
is spanned by $k$ even elements
$$a_i = \l(\begin{array}{cc}z^{i}&0\\0&z^{i}\end{array}\r), \qquad
i=0,1,\ldots,k-1,$$
and $k$ odd elements
$$\eta_i = \l(\begin{array}{cc}0&z^i\\-z^{n-2k+i}&0\end{array}\r), \qquad i=0,1,\ldots,k-1.$$
The OPE algebra is simply given by matrix multiplication, modulo the image of $\cD$. We can describe this algebra more compactly by
giving its generators and relations. There is one even generator $a=a_1$ and one odd generator $\eta=\eta_0$, and the relations are
\begin{equation}
\eta a=a\eta,\quad a^k=0,\quad \eta^2=-a^{n-2k}.\label{eq:rel_A}
\end{equation}
Note that for $k\leq n/3$ the second relation is equivalent to $\eta^2=0$. 

Morphisms between two different branes, $E_i$ and $E_j$, can also be calculated without difficulty, as explained in Ref.~\cite{ustwo}.
Here we merely quote the results.
As explained above, one may assume $i,j\le n/2$. For $i\le j$, the space of morphisms from 
$E_i$ to $E_j$ is spanned by $i$ even elements
\begin{equation*}
	a_k = \l(\begin{array}{cc}z^{j-i+k}&0\\0&z^{k}\end{array}\r), \qquad k=0,1,\ldots,i-1\no
\end{equation*}
and $i$ odd elements
\begin{equation*}
	\eta_k = \l(\begin{array}{cc}0&z^{k}\\-z^{n-i-j+k}&0\end{array}\r), \qquad k=0,1,\ldots,i-1.\no
\end{equation*}
The results for the case $i>j$ are obtained from these by parity reversal. Composition of morphisms is given by matrix multiplication, modulo
the image of $\cD$. 

In addition to the branes $E_k,$ $k\leq n/2$, we also have branes $E_k$, $k\geq n/2$. Since $E_{n-k}$ is the anti-brane of $E_k$,
no separate analysis of these branes is required. Note that the brane $E_k$ is not isomorphic to its anti-brane $E_{n-k}$.
Indeed, morphisms from $E_k$ to $E_{n-k}$ are the same as morphisms from $E_k$ to $E_k$, except for parity reversal. If $E_k$
were isomorphic to $E_{n-k}$, then there would be an invertible even morphism between them, or equivalently, there would be
an invertible odd morphism from $E_k$ to itself. One can easily see that there are no such morphisms (all odd endomorphisms are
nilpotent). The case $k=n/2$ is an exception, because the odd endomorphism $\eta$ is invertible, its inverse being $-\eta$. 
Thus the brane $E_{n/2}$ is isomorphic to its own anti-brane. We will express these facts by saying that the branes $E_k$, $k< n/2$
are orientable, while $E_{n/2}$ is unorientable.

It is shown in Ref.~\cite{Orlov} (see also comments in Ref.~\cite{ustwo}) that any topological D-brane is isomorphic to a sum
of the branes $E_k$ for some $k$. Thus we have a complete classification of B-branes in the $A$-type minimal models.

Before moving on, we take a closer look at 
a special example: the ``fundamental'' brane $E_1$. In this simplest case the algebra of endomorphisms is two-dimensional and spanned by
\begin{equation*}
	1 = \l(\begin{array}{cc}1&0\\0&1\end{array}\r), \qquad 
	\eta = \l(\begin{array}{cc}0&1\\-z^{n-2}&0\end{array}\r).\no
\end{equation*}
The only non-trivial algebraic relation of the OPE is
\begin{equation*}
	\eta\cdot\eta \;=\; \l\{ \begin{array}{ll}
	-1, \qquad & n=2\\
	0, & n>2\end{array}\r.\no
\end{equation*}
For $n>2$, the boundary OPE algebra is simply the exterior algebra $\wedge^*(V)$, where $V$ is a one-dimensional vector space. 
On the other hand, the OPE algebra for $n=2$ is the Clifford algebra $\Cl(1,\CC)$. The case 
$n=2$ is special because the LG model is massive, and the appearance of Clifford algebras in the boundary OPE is a generic feature of massive LG 
theories \cite{usone}.

Next we move on to the case $W=x^n-y^2$. There are two main differences compared to the case studied above. First, $\O$ is now the 
algebra of polynomials in two variables $x$ and $y$. Second, irreducible CDG modules in general have rank four rather than two, so 
the relevant space of endomorphisms is $\Mat(4,\O)$. Barring a subtlety that arises when $n\in2\ZZ$, which we shall discuss in detail below, 
irreducible B-branes in this theory are represented by
\begin{equation}
E_{k,\al}\;=\;\l\{\O^2_+\oplus\O^2_-, \;\;Q = \l(\begin{array}{cc}0&d_1\\d_0&0\end{array}\r)\r\}\label{eq:Ek_A2}
\end{equation}
where the ``$\pm$'' signs denote the $\ZZ_2$ grading, and 
$$d_0 = \l(\begin{array}{cc}x^k&\al\\-\be&-x^{n-k}\end{array}\r), \quad
	d_1 = \l(\begin{array}{cc}x^{n-k}&\al\\ -\be&-x^k\end{array}\r), \quad \al\be=y^2.$$
{\rm A priori}, $k$ runs from $0$ to $n$, and $\al\in\{1,y,y^2\}$.\footnote{Although there is a sign ambiguity for $\al$, 
one can change the sign by conjugating $d_0$ and $d_1$ with an appropriate matrix, therefore changing the sign from plus to minus
gives an isomorphic CDG module.}  As before, the localization 
principle tells us that the cases $k=0,n$ or $\al=1,y^2$ give trivial branes. Alternatively, one can check the triviality of these 
objects by explicitly computing their spaces of open string states using the algorithm described below. From this 
point on, we shall assume $\al=y$ and $k\neq0,n$. Furthermore, it is also clear that $E_k$ and $E_{n-k}$ define the same object in 
the category of B-branes, since they 
are related to each other by an automorphism which preserves the $\ZZ_2$ grading. The independent parameter range for irreducible branes can thus
be limited to $k=1,\ldots,[n/2]$. In fact, if $n\in2\ZZ$, the object $E_k,k=n/2$ is reducible, as discussed below, so we restrict ourselves to
$$k\;=\; 1,2,\ldots, \l[\frac{n-1}2\r], \qquad \al=y.$$

To compute the OPE, one notes that the boundary BRST operator acts on the space of endomorphisms, which is isomorphic to $\Mat(4,\CC[x,y])$, as 
follows
$$\D: \; \l(\begin{array}{cc}A&B\\C&D\end{array}\r)\mapsto\l(\begin{array}{cc}
	B  d_0 + d_1  C&\quad -A  d_1+d_1  D\\ -D  d_0+d_0  A& \quad C  d_1+d_0  B\end{array}\r)$$
where $A,B,C,D$ are $2\times2$ matrices with values in $\CC[x,y]$. The $\D$-closedness amounts to two independent conditions
$$WA=d_1D d_0, \qquad WC=-d_0Bd_0$$
which enable one to solve for $A$ and $C$ in terms of $B$ and $D$. After modding out $\D$-exact elements, we can describe
$\D$-cohomology using $B$ and $D$ only. The results are as follows. Even elements (bosonic open string states) are 
given by 
$$D \in \End(\O^2_-)\big/ \l(\im d_0^L\oplus\im d_1^R\r)$$
which satisfy the divisibility condition
$$W\,\big| \;d_1D d_0.$$
Here the superscripts on $d_0$ and $d_1$ indicates whether the operator acts from left or right. Similarly, odd elements (fermionic
open string states) are given by
$$B \in \Hom\l(\O^2_-,\O^2_+\r)\big/ \l(\im d_1^R\oplus\im d_1^L\r)$$
which satisfy
$$W\,\big| \;d_0Bd_0.$$

Let us first look at the bosonic sector. For the brane $E_k$, one can show that even elements of the endomorphism algebra can be 
represented in the 
quotient by 
$$D\in\begin{pmatrix}\CC[x]/x^k & \CC[x,y]/(x^n-y^2)\\\CC[x]/x^{n-k}&\CC[x]/x^{n-k}\end{pmatrix}.$$
The divisibility condition imposes a further relation
$$x^n-y^2\,\big|\;y(D_{22}-D_{11})-x^kD_{21}+x^{n-k}D_{12}.$$
This gives $2k$-dimensional even subspace, spanned by the following elements in $\End(\O^2_-)$:
$$a_i = \bmat x^i & 0\\0&x^i\emat,\quad a_{k+i} = \bmat 0&x^i\\x^{n-2k+i}&0\emat,\quad i=0,1,\ldots,k-1.$$

Similarly, one can show that odd elements live in the quotient 
$$B\in\begin{pmatrix}\CC[x,y]/(x^n-y^2) & \CC[x,y]/x^k\\\CC[x]/x^{k}&\CC[x]/x^{k}\end{pmatrix}\subset \Hom\l(\O^2_-,\O^2_+\r)$$
and satisfy the relation
$$x^n-y^2\,\big|\;x^{n-k}B_{22}-x^kB_{11}+y(B_{12}-B_{21}).$$
Therefore the odd subspace of the boundary OPE algebra is spanned by the following elements in $\Hom(\O^2_-,\O_+^2)$:
$$\eta_i = \bmat 0&x^i\\x^i&0\emat,\quad \eta_{k+i} = \bmat x^{n-2k+i}&0\\0&x^i\emat,\quad i=0,1,\ldots,k-1.$$
The full boundary OPE algebra can be described by generators and relations as follows. It has one even generator $a=a_1$, which is central, 
and two odd generators $\xi=\eta_0$ and $\eta=\eta_k$, which satisfy the relations
\begin{equation}
a^k=0,\quad \xi^2=1, \quad \eta^2 = -a^{n-2k}, \quad \xi\eta+\eta\xi=0.\label{eq:rel_A2}
\end{equation}
Obviously this algebra does not decompose as a direct sum, so the branes $E_k$ are all irreducible. We also see that for all $k$ there
is an invertible odd endomorphism $\xi$. As discussed above, this means that the branes $E_k$ are unorientable (are isomorphic to
their own anti-branes). 

The case $n\in2\ZZ$ and $k=n/2$ is somewhat special in that the object $E_{n/2}$ is reducible. In fact, it splits into two irreducible 
branes corresponding to the 
following CDG modules:
$$E_{\pm}=\l\{\O_+\oplus\O_-, \;\;Q_\pm = \bmat 0&x^{n/2}\pm y\\x^{n/2}\mp y&0\emat\r\}$$
Clearly $E_-$ is the anti-brane of $E_+$. 

It is a simple matter to determine the boundary OPE algebras. The endomorphism algebra of $E_\pm$ is spanned by $n/2$ even elements:
\begin{equation*}
	a_i = \l(\begin{array}{cc}x^i&0\\0&x^i\end{array}\r), \qquad i=0,1,\ldots,\frac{n}{2}-1.\no 
\end{equation*}
In contrast to branes $E_k$ discussed above, there are no odd elements. This algebra has one generator $a=a_1$ and a relation $a^{n/2}=0$.
One can likewise work out the space of boundary changing operators living in $\Hom(E_+,E_-)$. It is purely odd and has dimension $n/2$.
This implies that the branes $E_+$ and $E_-$ are not isomorphic, and therefore both are orientable. 

Note that in all previous examples the Witten index of the space of topological strings from a brane to itself was zero.
For the branes $E_\pm$ the Witten index is equal to $n/2$.

\subsection{Orbifolded $A$-type minimal models}
\label{subsec:A2}

For any Landau-Ginzburg model the category of D-branes has an obvious symmetry (autoequivalence): parity-reversal, which takes
branes to their anti-branes. We may consider orbifolding the category by this symmetry. Note that this symmetry does not
act on the target space of the LG model, but only on the category of D-branes. Such symmetries are often called {\it quantum} symmetries.
{}From the string theory point of view, these are symmetries of the world-sheet theory which do not originate from any symmetry
of the target space. A well-known example of a quantum
symmetry is T-duality, which, in its most basic form, states that the quantum sigma-model whose target is a torus with a flat
metric is unchanged if one replaces the torus with its dual. The mathematical counterpart of this phenomenon is
the Fourier-Mukai transform which identifies the derived categories of an abelian variety and its dual. The situation in our case
is similar, but simpler. The physical counterpart of parity-reversal is a non-geometric $\ZZ_2$ symmetry of 
the SCFT which acts trivially on the
NS-sector states and by $-1$ on all RR-sector states (there are no mixed NS-R states in our case). Since chiral primary states
reside in the NS sector, the chiral rings of the original and orbifolded theories are identical. In other words,
orbifolding has no effect on topological closed-string states. But the properties of Cardy branes in the two theories
are different, as was shown in Ref.~\cite{MMS}. Therefore we expect that the orbifolded category of topological branes
is different from the unorbifolded one. We will see that this is indeed the case for $W=z^n$. Moreover, we will see that topological
D-branes in the orbifolded category are the same as topological D-branes in the LG model $W=z^n+y^2$. This lends credence
to our conjecture that adding a square to $W$ has the same effect on the category of B-branes as orbifolding by parity-reversal.

The construction of the orbifolded category was explained in the end of Ref.~\cite{ustwo}. Rephrasing this construction a little, we
get the following. Objects are pairs $(E,\beta)$, where $E$ is a D-brane in the original (unorbifolded) theory, and $\beta$ is
an odd endomorphism of $E$ satisfying $\beta^2=1$. Morphisms (boundary-changing operators) between $(E,\beta)$ and $(E',\beta')$ are
morphisms from $E$ to $E'$ intertwining $\beta$ and $\beta'$:
$$
\phi\beta =(-1)^{|\phi|}\beta'\phi,\quad \phi\in \Hom(E,E').
$$
Note that given an object $(E,\beta)$ we have another valid object $(E,-\beta)$. This other object may or may not be isomorphic
to $(E,\beta)$. We will see examples of both possibilities below.

Let us apply this construction to D-branes in the $A$-type minimal model $W=z^n$. Clearly, for $\beta$ to exist, $E$ must be isomorphic
to its own anti-brane. Irreducible branes $E_k$ in the model $W=z^n$ are not their own anti-branes, except in the case $k=n/2$.
Thus we are forced to consider direct sums $F_k=E_k\op \bE_k=E_k\op E_{n-k}$. Clearly, it is sufficient to take $k$ in the range
$0<k<n/2$ (the case $k=n/2$ is special and will be considered separately). The endomorphism algebra of $F_k$ can be inferred from the
known answer for the endomorphism algebra of $E_k$, and we see that the only invertible odd endomorphism is, up to a scalar multiple, 
$$
\beta=\begin{pmatrix} 0 & 1 \\ 1 & 0\end{pmatrix}
$$
where we write $\beta\in \End(E_k\op\bE_k)$ as a block matrix with elements in $\End(E_k)$, $\Hom(\bE_k,E_k),$ $\Hom(E_k,\bE_k),$
and $\End(\bE_k)$. The scalar multiple is fixed to be $\pm 1$ by the requirement $\beta^2=1$. Finally, it is easy to see that the branes
corresponding to the two possible signs are actually isomorphic, an isomorphism being 
$$
\phi=\begin{pmatrix} 1 & 0 \\ 0 & -1\end{pmatrix}
$$
in the same notation. The endomorphism algebra of $F_k$ in the orbifolded category consists of endomorphisms of $E_k\op\bE_k$ in the
unorbifolded category which supercommute with $\beta$, i.e. have the form
$$
\phi=\begin{pmatrix} A & B \\ (-1)^{|B|} B & (-1)^{|A|} A\end{pmatrix}.
$$
Here $A$ is an arbitrary element of $\End(E_k)$, $B$ is an arbitrary element of $\Hom(\bE_k,E_k)$, and $|A|$ and $|B|$ denote
the parities of $A$ and $B$.

It is easy to check that the number and endomorphism algebras of the branes $F_k$ agree with the number and endomorphism algebras of
the rank-four branes in the LG model $W=x^n-y^2$. This confirms our conjecture that orbifolding by parity-reversal is equivalent
to adding a square to the superpotential.

It remains to consider the case $k=n/2$. The brane $E_{n/2}$ in the model $W=z^n$ is isomorphic to its own anti-brane, so all we have
to do to complete it to an object of the orbifolded category is to find a suitable $\beta$. From the results of the previous subsection
we know that up to a scalar multiple there is only one invertible odd endomorphism of $E_{n/2}$, which we denoted $\eta$.
The requirement $\beta^2=1$ tells us that $\beta=\pm i\eta$, so we get two possible objects $F_+=(E_{n/2},i\eta)$ and
$F_-=(E_{n/2},-i\eta).$ Unlike in the previous case, these two objects are {\it not} isomorphic. Indeed, the space of morphisms
from $F_+$ to $F_-$ consists of endomorphisms of $E_{n/2}$ which anti-super-commute with $\eta$. All such endomorphisms are odd, and so
$F_+$ cannot be isomorphic to $F_-$. Instead, there is an odd invertible morphism from $F_+$ to $F_-$ given by $\eta$ itself,
and this means that $F_-$ is the anti-brane of $F_+$. The endomorphism algebra of $F_+$ consists of endomorphisms of $E_{n/2}$ which 
super-commute with $\eta$. It is easy to see that the space of such endomorphism is purely even and spanned by $a^p$, $p=0,\ldots,n/2-1$
in the notation of the previous subsection. It is also easy to see that the properties of $F_\pm$ match the properties of the
branes $E_\pm$ in the model $W=x^n-y^2$.

\subsection{$D$-type minimal models}
\label{subsec:D}

The $D$-type minimal models have the following LG realization:
$$W_{D_{n+2}}\;=\;x^{n+1}-xy^2.$$ As always, one may add another free massive chiral field $z$ to the theory, modifying the superpotential by
 $z^2$. We will comment on the effect of this at the end of this section. For the moment, we focus on the above superpotential. The most obvious
way to factorize $W$ gives the following brane:
\begin{equation*}\label{obvbrane}
E\;=\;\l\{\O_+\oplus\O_-, \;Q = \bmat0&x\\x^{n}-y^2&0\emat\r\}
\end{equation*}
Swapping $x$ and $x^n-y^2$ gives the anti-brane of ${E}$. The endomorphism algebra of $E$ is spanned by two even elements
\begin{equation*}
	a_0 = \l(\begin{array}{cc}1&0\\0&1\end{array}\r), \qquad 
	a_1 = \l(\begin{array}{cc}y&0\\0&y\end{array}\r).\no
\end{equation*}
The element $a_0$ is the identity, while $a_1$ satisfies $a_1^2=0$ in the $\cD$-cohomology.
Thus the boundary OPE algebra is isomorphic to the exterior algebra $\wedge^*(\CC)$ as an ungraded algebra. The anti-brane $\bE$ has the
same endomorphism algebra, while the space of morphism from $E$ to $\bE$ is two-dimensional and purely odd. This implies that $E$ is
not isomorphic to $\bE$, and therefore the brane $E$ is orientable.

For even $n$, there are two additional branes given by
\begin{eqnarray*}
	E_1&=&\l\{\O_+\oplus\O_-, \;Q = \bmat0&x(x^{n/2}-y)\\x^{n/2}+y&0\emat\r\},\no\\
	E_2&=&\l\{\O_+\oplus\O_-, \;Q = \bmat0&x(x^{n/2}+y)\\x^{n/2}-y&0\emat\r\}.\no
\end{eqnarray*}
The space of endomorphisms of
$E_1$ is purely even and spanned by the following $n/2+1$ elements:
$$a_i = \l(\begin{array}{cc}x^i&0\\0&x^i\end{array}\r), \qquad i=0,1,\ldots,\frac{n}{2}.$$
The boundary OPE algebra is is generated by $a=a_1$ with the relation $a^{n/2}=0$. The brane $E_2$ has the same 
OPE algebra. It is easy to see that $\bE_i$ is not isomorphic to $E_i$, and therefore both $E_1$ and $E_2$ are orientable.

Since the OPE algebras of $E_1$ and $E_2$ are identical, one naturally asks whether they define the same B-brane. To answer this question one 
needs to work out the space of boundary changing operators. By using the now hopefully familiar algorithm, one can show that the 
space of boundary 
changing operators living in $\Hom(E_1,E_2)$ is purely odd and is spanned by
$$\eta_{i} = \l(\begin{array}{cc}0&-x^{i+1}\\x^i&0\end{array}\r), \qquad i=0,1,\ldots,\frac{n}{2}-1.$$
Since the dimension of this space is one less than the dimension of the space of endomorphisms on $E_1$, 
one concludes that $E_2$ is isomorphic neither to $E_1$ 
nor to the anti-brane of $E_1$. Therefore $E_1$ and $E_2$ define two different irreducible B-branes.

There are other branes which are not included in the above analysis. In order to see these extra branes, one needs to consider CDG modules of 
rank four. Specifically, we consider the following CDG modules:
$$ E_{k,\al} \;=\; \l\{\O^2_+\oplus\O^2_-, \;\;Q = \bmat0&d_1\\d_0&0\emat\r\}$$
where
$$d_0 = \bmat x^k&\al\\-\be&-x^{n+1-k}\emat, \quad
	d_1 = \bmat x^{n+1-k}&\al\\ -\be & -x^k\emat,\quad \al\be = xy^2.$$
Apart from a few slight differences, the situation here is very much like in the case $W=x^n-y^2$ analyzed above. As before, the cases of 
$k=0, n+1$ or $\al=\pm 1,\pm xy^2$ give zero objects, so we exclude them from our list. 
The first difference compared to the case $W=x^n-y^2$ comes 
from the fact that replacing either $k\lr n+1-k$ or $\al\lr\be$ does not give the same brane, but the anti-brane. In other words, 
all these branes are orientable while 
in the case $W=x^n-y^2$ they are unorientable. The second difference is that there appear to be more types of branes. Naively, there seems
to be many different ways to choose $\alpha$ and $\beta$ so that $\alpha\beta=xy^2$. However, changing the signs of both $\alpha$ and 
$\beta$ can be undone by an automorphism which does not change the grading.
It is also easy to see that exchanging $\alpha$ and $\beta$ is equivalent to replacing a brane with its anti-brane. Thus
we can restrict to the following range:
$$k=1,2,\ldots,\l[\frac{n+1}2\r] \qquad \be= x, y.$$
These branes, together with their anti-branes, exhaust all the objects $E_{k,\al}$ defined above.

The rest of the calculation is essentially the same as before, and we summarize the results below.

\subsubsection*{The case $\be=x$.} The even generators of the space of morphisms live in the quotient space 
$\End(\O_-^2)\big/ \l(\im d_0^L\oplus\im d_1^R\r)$. They can be represented by the following matrices:
\begin{equation}
\l(\begin{array}{cc}a&\mu\\\nu&b\end{array}\r), \quad a,b\,\in\CC[y]/(y^2), \; \nu\in\CC[y],\; \mu\in\CC[x,y]/(x^n-y^2).\label{eq:mat}
\end{equation}
The divisibility condition leads to two independent relations:
$$x^n-y^2\,\big|\;x^{k-1}(a-b)-\mu+x^{2k-2}\nu, \qquad x\,|\,\nu.$$
It follows from these relations that the even part of the endomorphism algebra is spanned by the following elements in $\End(\O^2_-)$:
$$a_1=\bmat 1&0\\0&1\emat, \; a_2 = y\cdot a_1, \; a_3 = \bmat 1 & x^{k-1}\\0&0\emat, \; a_4 = y\cdot a_3.$$

The odd part of $\cD$-cohomology can also be parametrized by matrices of 
the form of Eq.~(\ref{eq:mat}). The divisibility condition now implies
$$x^n-y^2\,\big|\;x^{k-1}a-x^{n-k}b-\mu, \qquad \nu=0.$$
Thus the odd subspace is four-dimensional and is spanned by
$$\eta_1=\bmat 1&x^{k-1}\\0&0\emat, \; \eta_2 = y\cdot\eta_1, \; \eta_3 = \bmat 0 & -x^{n-k}\\0&1\emat, \; \eta_4 = y\cdot\eta_3.$$

The resulting algebra of endomorphisms can be described by generators and relations. There are two even generators $a=a_2$ and $b=a_3$,
two odd generators $\eta_1$ and $\eta_3,$ and the following relations:
\begin{gather*}
a^2=0, \quad b^2=b, \quad \eta_1^2=0,\quad \eta_3^2=0, \quad \eta_1\eta_3+\eta_3\eta_1=-1,\\
b\eta_1=0,\quad b\eta_3=\eta_3.
\end{gather*}
In addition, $a$ and $b$ commute with everything.
It is not hard to see that this algebra is isomorphic to the endomorphism algebra of the object $E\op \bE$, where $E$ is the
brane given by Eq.~(\ref{obvbrane}). This suggests that the brane we are considering is isomorphic to
the direct sum of the brane Eq.~(\ref{obvbrane}) and its anti-brane. To verify that two objects are isomorphic, one has to compute the spaces of
morphisms between them (in both directions) and check that there exists a pair of even morphisms, going in opposite directions,
whose compositions (in either direction) are the identity
endomorphisms. We have checked that this is indeed the case. This example illustrates that two very different CDG modules
can become isomorphic upon passing to the derived category. In physical terms, very different tachyon profiles can produce the same
topological brane.

\subsubsection*{The case $\be=y$.} The even subspace can be represented by matrices of the form
$$D\in\bmat \CC[x]/x^k & \CC[x,y]/(x^{n+1}-xy^2)\\ \CC[x]/x^{n+1-k} & \CC[x]/x^{n+1-k}\emat\subset \End(\O^2_-)$$ that are subject to the 
divisibility condition
$$x^{n+1}-xy^2\,\big|\;yx^{k}(D_{22}-D_{11})+y^2D_{12}-x^{2k}D_{21}.$$
Solving this algebraic relation gives $2k$ even elements spanning the even subspace of $\cD$-cohomology:
$$a_i=\bmat x^i & 0\\0&x^i\emat, \quad 
a_{k+i} = \bmat 0& x^{i+1}\\x^{n-2k+1+i}&0\emat, \quad i=0,1,\ldots,k-1.
$$

The odd subspace is computed in the same way and is spanned by the following elements in $\Hom(\O^2_-,\O^2_+)$:
$$\eta_i = \bmat x^{n-2k+1+i} & 0\\0&x^i\emat, \quad 
\eta_{k+i} = \bmat 0& x^{1+i}\\x^{i}&0\emat, \quad i=0,1,\ldots,k-1.$$

The endomorphism algebra of the brane has two odd generators $\xi=\eta_0$ and $\eta=\eta_k$. The even element $a_i$ is expressed as $\eta^{2i}$.
The relations are
$$
\eta^{2k}=0, \quad \xi^2 = -\eta^{2n-4k+2}, \quad \xi\eta+\eta\xi=0.
$$
Note that the Witten index vanishes for all $k$.
It is also easy to see that all these branes are irreducible, orientable, and pairwise non-isomorphic. 

We conjecture that the branes constructed above are the only irreducible branes in $D$-type minimal models. We expect that this can be
proved using the results of Ref.~\cite{Orlov}.

As in the case of $A$-type minimal models, we can consider orbifolding by the $\ZZ_2$ symmetry which exchanges branes and anti-branes.
We expect that the effect of this is the same as adding a square to the superpotential.

\subsection{$E$-type minimal models}

Let us list a few simple examples of branes in $E$-type minimal models. Our list is not supposed to be exhaustive.

The $E_6$ minimal model has a \LG realization with 
$$
W_{E_6} \;=\; x^3+y^4.
$$
Examples of irreducible B-branes are given by 
$$
E_{k,l}=\l\{\O^2_+\oplus\O^2_-, \;\;Q = \bmat 0&d_1\\d_0&0\emat\r\}
$$
where
$$
d_0 = \l(\begin{array}{cc}x^k&y^{\ell}\\y^{4-\ell}&-x^{3-k}\end{array}\r), \quad
d_1 = \l(\begin{array}{cc}x^{3-k}&y^{\ell}\\ y^{4-\ell}&-x^k\end{array}\r).
$$
One may choose the fundamental range to be $k=1$ and  $\ell\in\{1,2\}$. The brane $E_{1,1}$ is orientable, and its anti-brane is $E_{1,3}$. 
On the other hand, $E_{1,2}$ is unorientable. 

The boundary OPE algebras for branes in the fundamental range are as follows. There is a single even generator $y$ which is central. 
There are also two odd generators $\xi$ and $\eta$. The boundary OPE algebra is specified by the following relations

\bea
y^{\ell} &=& \eta^2 \;=\; 0,\no\\
\label{eq:E6}\xi^2 &=& -y^{4-2\ell},\\
\xi\eta &=& -\eta\xi. \no
\eea

The $E_8$ minimal model has a \LG realization with 
$$
W_{E_8} \;=\; x^3+y^5.
$$
{F}or our purposes, this is very similar to the $E_6$ case analyzed above. In particular, one immediately concludes that there are two 
irreducible B-branes labeled by $E_{1,1}, E_{1,2}$. 
The main difference here is that both of them are orientable now, their anti-branes being given by $E_{1,4}$ and $E_{1,3}$ respectively. 
The boundary OPE is the same as given in Eq.~(\ref{eq:E6}), except that $y^4$ must be replaced with $y^5$ everywhere. Furthermore, in 
the case of $E_{1,2}$, the even generator $y$ is no longer an independent generator (because $y=-\xi^2$), and the boundary OPE algebra 
is generated by two odd generators.

The case of $E_7$ is slightly different. The \LG superpotential is
$$
W_{E_7} \;=\; x^3+xy^3.
$$
First of all, there is a brane defined by 
$$
E=\l\{\O_+\oplus\O_-, \;\;Q = \bmat 0&x^2+y^3\\x&0\emat\r\}.
$$
This brane is orientable, its anti-brane having the transposed $Q$. In addition, there are branes specified by higher rank objects:
$$ 
E_{\al} \;=\; \l\{\O^2_+\oplus\O^2_-, \;\;Q = \bmat0&d_1\\d_0&0\emat\r\}
$$
where
$$
d_0 = \bmat x&\al\\\be&x^2\emat, \quad
d_1 = \bmat x^2&\al\\ \be & -x\emat,\quad \al\be = xy^3.
$$
{\em A priori}, there are three independent branes associated with the choices $\be = x,y, y^2$. All of them appear orientable, and their 
anti-branes are obtained by swapping $\al$ and $\be$. However, as in the $D$-type case, the object with $\be=x$ is decomposable. 
In fact one can show that it is isomorphic to the direct sum $E\op\bar{E}$, and therefore is unorientable.
The objects associated with $\be=y$ and $\be=y^2$, on 
the other hand, are irreducible and orientable. We leave the details to the reader as an exercise.

\subsection{Disk correlators in topological minimal models}

We will now apply the general formula derived in Ref.~\cite{ustwo} for topological open string correlators to the branes obtained above. 
For simplicity, we focus on disk correlators but generalization to higher genera and multi-boundary cases is straightforward. 
Up to an unessential numerical factor, the disk correlator with a bulk insertion 
$\al\in\O/\part W$ and a boundary insertion $\phi$ is given by
\begin{equation}
\big\langle\al\cdot\phi\big\rangle_{\rm disk} \;=\; \frac1{n!\,(2\pi i)^n} \oint_L \frac{\alpha\cdot\str\big[(\partial Q)^{\wedge n}\phi\big]}
{\part_1W\part_2W\ldots\part_nW},\label{eq:disk}
\end{equation}
where the integration is carried out over an $n$-dimensional Lagrangian torus. For a more detailed explanation of the formula, 
see Section~\ref{sec:intro} or Ref.~\cite{ustwo}.

Let us consider an $A$-type minimal model whose \LG realization has $W=z^n$.  
The Jacobi ring is $\CC[z]/z^{n-1}$, and it suffices to take the bulk operator to be $\al=z^i$ with $i\le n-2$. 
As shown in subsection \ref{subsec:A}, irreducible B-branes are labeled by $E_k$, and it is sufficient to restrict
the range of $k$ to $0<k\le n/2$.  The boundary operator algebra of $E_k$ has an even generator $a$ and an odd generator $\eta$, 
with relations given in Eq.~(\ref{eq:rel_A}). One easily sees that the supertrace $\str[\part Q\cdot\phi]$ vanishes unless 
$\phi=a^\ell\eta$. In particular, all disk correlators with no boundary insertions vanish. From the closed string point of view, 
this means that they carry no RR charge. We will come back to this issue when we make comparison with the boundary state formalism in Section~\ref{sec:cardy}.

When the boundary operator {\em is} given by $\phi=a^\ell\eta$, our general formula Eq.~(\ref{eq:disk}) yields the following result:
$$\big\langle z^i\cdot a^\ell\eta\big\rangle_{E_k} \;=\; \Big\{\begin{array}{rl}
	-1 & \qquad {\rm if}\;\; \ell+i=k-1,\\
	0 & \qquad \quad {\rm otherwise.}\end{array}$$

As explained in sections \ref{subsec:A} and \ref{subsec:A2}, B-branes in the orbifolded $A$-type minimal models are the same as B-branes in 
LG theories with $W=x^n-y^2$. In order to compute disk correlators, we will use the latter description.
The usual irreducible B-branes are labeled by $E_{k,\al}$ with $0<k\le[\frac{n-1}2]$ and $\al=y$, as listed in (\ref{eq:Ek_A2}). 
In case of even $n$, there are two additional irreducibles $E_\pm$. Let us consider the branes $E_{k,y}$ first. 
The ring of bulk operators is still given by the Jacobi ring $\CC[x]/x^{n-1}$. The boundary operator algebras are 
given in Eq.~(\ref{eq:rel_A2}). An easy computation shows that the supertrace vanishes unless the boundary operator $\xi\eta$ is inserted, 
so it suffices to consider insertions of the form $x^i\cdot a^\ell\xi\eta$. The disk correlators are computed by the integral
$$
\frac1{(2\pi i)^2}\oint_{L}\frac{-2nx^{n+\ell+i-k-1}\,dx\wedge dy}{\part_1W\part_2W},
$$
which gives
$$
\big\langle x^i\cdot a^\ell\xi\eta\big\rangle_{E_{k,y}} \;=\; 
\Big\{\begin{array}{ll} 1 & \qquad {\rm if}\;\; \ell+i=k-1\\
	0 & \qquad \quad {\rm otherwise}\end{array}
$$
Like the branes in the case $W=z^n$ discussed above, these branes carry no RR charge.

Finally let us consider the branes $E_\pm$ which exist when $n$ is even.  
In the case of $E_+$, there is a single even generator 
$a$ with a relation $a^{n/2}=0$. One can show that the disk correlator for a general insertion $x^i\cdot a^\ell$ is
\bea
	\big\langle x^i\cdot a^\ell\big\rangle_{E_+} &=& 
	\frac1{(2\pi i)^2}\oint_{L}\frac{-nx^{n/2+\ell+i-1}\,dx\wedge dy}{\part_1W\part_2W}\no\\
	&=& \Big\{\begin{array}{cl}
	1/2 & \qquad {\rm if}\;\; \ell+i=n/2-1\\
	0 & \qquad\quad {\rm otherwise.}\end{array}\no
\eea
The disk correlators for $E_-$ are obtained by a simple sign flip. Notice that in the particular case $\ell=0$ (no boundary insertion), 
the nonvanishing disk correlators are
$$
\langle x^{n/2-1}\rangle_{E_+} = - \langle x^{n/2-1}\rangle_{E_-} = 1/2.
$$ 
In this special case disk correlators have also been computed in Ref.~\cite{Hbook}. Since not all disk correlators without
boundary insertions vanish, the branes $E_\pm$ carry nonzero RR charge. There are also RR-charged branes in $D$-type minimal models,
as discussed in the Appendix.

Recall that in the case of even $n$, the object $E_{n/2,y}$ is reducible and is isomorphic to the direct sum $E_+\oplus E_-$. 
The disk correlators computed above are compatible with this assertion. Even though both $E_+$ and $E_-$ are RR-charged, 
their direct sum carries no RR charge as $\langle x^i\rangle_{E_+\op E_-} = 0$ for all $i$. This is in agreement with our
earlier statement that $E_{n/2,y}$ is RR-neutral.

This exhausts our list of B-branes in topological $A$-type minimal models and their $\ZZ_2$ orbifolds. 
One can likewise compute disk correlators for B-branes in $D$-type and $E$-type minimal models. 
We give explicit results for the $D$-type minimal models in the Appendix and leave the $E$-type to an interested reader.

\section{Comparison with the boundary state formalism}
\label{sec:cardy}

In this section we compare our classification of B-branes in $\N=2$ minimal models with the known results obtained from the boundary state 
formalism. Since, to the best of our knowledge, Cardy branes in $D$-type $\N=2$ minimal models have not been studied in the literature, 
we shall limit the discussion to $A$-type minimal models.

\subsection{General remarks}

At the outset, we should address one point which may be puzzling at first sight. It seems reasonable to assume that adding squares to the 
superpotential
should not change the low-energy physics, including the topological observables. This is quite obvious in the closed-string sector,
because the Jacobi ring is unaffected by the addition of squares. But the situation in the open-string sector is more complicated. It has been
shown in Ref.~\cite{Orlov} that the spectrum of topological branes, as well as the boundary OPE algebra, are unaffected by the addition of
{\it two} squares to $W$. But we have seen in subsection~\ref{subsec:A} that adding a {\it single} square to $W=z^n$ has a non-trivial effect 
on the category of D-branes.
In subsection~\ref{subsec:A2} we have shown that the effect of adding a single square is the same as the effect of orbifolding the category by 
a $\ZZ_2$ symmetry
which exchanges branes and anti-branes (at least, in this special case, and probably in general). On the other hand, it is believed
that given a superconformal field theory the spectrum of D-branes is uniquely determined. In particular, there should be a unique answer
to the question ``What is the spectrum of Cardy branes in an $\N=2$ minimal model?'' 

This apparent conflict between the Landau-Ginzburg 
and Cardy approaches is resolved by the observation that in fact for every ADE Dynkin
diagram there are {\it two} closely related SCFTs which have the same chiral ring. We will limit our to discussion to 
$A$-type Dynkin diagrams. The usual $A$-type minimal model corresponds to the diagonal 
combination of characters of the $\N=2$ super-Virasoro algebra. 
In the notation of Ref.~\cite{BH}, its Hilbert space is given by
$$
\cH=\sum_{(j,n,s)} \cH_{j,n,s}\ot \cH_{j,-n,-s}.
$$
Here the allowed values of the labels $j,n,s$ are
\begin{gather}
j\in \left\{0,\frac{1}{2},1,\ldots,\frac{k}{2}\right\},\quad n\in \ZZ/(2k+4)\ZZ,\quad s\in \ZZ/4\ZZ,\no\\
2j+n+s=0\ {\rm mod}\ 2.\no
\end{gather}
The summation is over all distinct values of $(j,n,s)$, taking into account the following equivalence relation:
$$
(j,n,s)\sim (k/2-j,n+k+2,s+2).
$$
The spaces
$$\op_{s\, {\rm even}} \cH_{j,n,s}$$
and
$$
\op_{s\, {\rm odd}} \cH_{j,n,s}
$$ 
are irreducible representation of the $\N=2$ super-Virasoro algebra with central charge $c=3k/(k+2)$. Even and odd values
of $s$ correspond to NS and R sectors, respectively. The variable $s$ is related to the eigenvalue of the fermion number
operator $F$ by 
$$e^{\frac{i\pi s}{2}}=e^{-\pi i F}.$$ 
Note that in accordance with this definition  $F$ has half-integral eigenvalues in the Ramond sector. The Witten index in the
Ramond sector is therefore defined as 
$$
\Tr_{s\, {\rm odd}} (-1)^{F+\frac12}.
$$
Note also that the above partition function corresponds to the non-chiral GSO projection of type 0A (i.e. opposite projection
in the left-moving and right-moving Ramond sectors).
We will call this theory the $k^{\rm th}$ minimal model (of type $A$) and denote it ${\rm MM}_k$.

The $\ZZ_2$ action of interest to us is diagonal in this
basis, with eigenvalue $e^{i\pi s}$. Orbifolding by this $\ZZ_2$ projects out all RR states, but all the twisted sector states are
again in the RR sector. The Hilbert space of the orbifolded theory is
$$
\cH'=\sum_{(j,n,s)} \cH_{j,n,s}\ot \cH_{j,-n,s}
$$
This partition function corresponds to the non-chiral GSO projection of type 0B (i.e. the same projection in the left-moving
and right-moving Ramond sectors). We will denote this theory ${\rm MM}_k/\ZZ_2$.
The two SCFTs are obviously different. But note that the NS-NS sectors in the two theories are the same, hence the chiral rings 
are also identical. In other words, on a world-sheet without boundaries orbifolding has no effect on topological 
correlators.\footnote{Note that although the two theories have the same chiral ring, the number of RR ground states is different.
In fact, there are no RR ground states in the Hilbert space $\cH'$. The usual spectral flow argument does not apply here, because
$\cH'$ is not invariant with respect to spectral flow by $\theta=1/2$. Despite this, the partition function of the orbifolded
theory is modular invariant,  as one can easily verify.}

On the other hand, properties of B-branes in the two models are rather different, even in the topological sector~\cite{MMS}.
For example, in the unorbifolded $n-2^{\rm nd}$ minimal model\footnote{Of course,
which of the two is the unorbifolded one is a matter of convention. Each of the two models has a $\ZZ_2$ symmetry, orbifolding
by which gives the other model.} with $n$ even there are $\frac{n}{2}-1$ unorientable irreducible branes with zero Witten index, and
two additional branes, which are both orientable, are anti-branes for each other, and have Witten index $n/2$. On the other hand,
in the orbifolded $n-2^{\rm nd}$ minimal model with $n$ even, there are $\frac{n}{2}-1$ orientable branes, their anti-branes, and one 
more unorientable brane; all these branes have zero Witten index. Comparing with the results of Section~\ref{sec:minimal}, 
we see that Cardy branes in the unorbifolded minimal
model seem to match topological branes in the LG model $W=x^n-y^2$, while Cardy branes in the orbifolded minimal model match those
in the LG model $W=z^n$. Below we will check this identification in more detail. The relation between branes in topological LG
models and Cardy B-branes in minimal model has been also discussed in Refs.~\cite{Hbook,BH} (in some special cases). Our results
are in agreement with these papers.

\subsection{Cardy B-branes in ${\rm MM}_k$}

The standard Cardy construction leads naturally to A-branes. In particular there is a one-to-one correspondence between RR ground states and 
Cardy A-branes. To obtain B-branes from a Cardy-like construction, one starts with Cardy A-brane boundary states 
in the orbifolded model ${\rm MM}'_k={\rm MM}_k/\ZZ_2\times\ZZ_{k+2}$, and identifies those which are invariant under 
$\ZZ_2\times\ZZ_{k+2}$ as B-brane boundary 
states in the original ${\rm MM}_k$. The idea  of this construction is that orbifolding by $\ZZ_2\times\ZZ_{k+2}$ acts as a mirror
symmetry in this case. The boundary states of these B-branes are constructed in Ref.~\cite{MMS}:
\begin{equation}\label{Bone}
|j,s\rangle_{\rm B} = \sqrt{2(k+2)}\sum_{2j',s' {\rm even}}
\frac{S_{j,-2j-s,s}^{j',0,s'}}{\sqrt{S_{0,0,0}^{j',0,s'}}}\,|{\rm B};j',0,s'\drangle,
\end{equation}
where $|{\rm B};j',0,s'\drangle$ are the Ishibashi states in ${\rm MM}'_k$. Independent Cardy states are parametrized by
$$2j=0,1,\ldots,\l[\frac{k}2\r], \qquad s=0,1.$$
The two choices of $s$ correspond to different B-type supersymmetric boundary conditions $\bar{G}_+=\pm\bar{G}_-$. We can choose $s=1$ if
we fix a sign convention. Note that these boundary states have only NS component. This means that the corresponding branes are
unorientable.

Our goal is to compute the spectrum of chiral primary states in the open-string NS sector. It is actually more convenient to
compute the number of Ramond ground states, which are obtained from chiral primary states by spectral flow. One can read off the
spectrum of Ramond ground states from the boundary state overlaps $_B\langle j_1,s_1|q_c^{L_0-c/24}|j_2,s_2\rangle_B.$
The even Ramond states can be read off
\bea
_{\rm B}\!\langle j,0|q_c^{L_0-c/24}|j,1\rangle_{\rm B} &=& \sum_{(\tj,\tn)}\l(N_{jj}^{\tj}+N_{jj}^{k/2-\tj}\r)\chi_{\tj,\tn,1}(q_o)\no
\eea
where
$$N_{ij}^{\ell} = \l\{\begin{array}{l}1, \quad \mbox{if $|i-j|\le \ell\le$min$\{i+j,k-i-j\}$,}\;\;i+j+\ell\in\ZZ\\
	0, \qquad \mbox{otherwise}\end{array}\r.$$
are the fusion coefficients of $\su(2)_k$. 

For our purpose, we are only interested in the multiplicity of ground states, i.e. $(\tj,\tn,\ts)=(\ell,2\ell+1,1)$. This has a simple form
\bea
n_{jj}^{\ell} &=& \l\{\begin{array}{l}1, \quad \ell\in\{0,1,\ldots, 2j\}\cup\{k,k-1,\ldots, k-2j\}\\0, \qquad\qquad {\rm otherwise}\end{array}\r.\no
\eea
It follows that there are in total $2(2j+1)$ bosonic chiral primary states. 

Odd Ramond states can be read off another overlap:
$$
_{\rm B}\!\langle j,0|q_c^{L_0-c/24}|j,-1\rangle_{\rm B}.
$$
In general, $s=1$ and $s=-1$ boundary states are almost the same, the only difference being the sign of the RR contribution.
Therefore in general the number of even and odd open-string states
is different. But in the present case, the boundary states have no RR piece, and therefore the $s=1$ and $s=-1$ boundary
states are identical. This implies that the number of odd and even open-string states is the same.

If $k\in2\ZZ$, there is an additional subtlety. The above boundary state $|k/4,s\rangle_{\rm }$ is not an irreducible 
brane, since the multiplicity of the vacuum representation is $n_{k/4,k/4}^0 = 2$. Instead it is a sum of two irreducible B-branes, 
denoted $|k/4,\pm1\rangle_{\rm \tilde{B}}$, which are anti-branes of each other. An explicit form of the boundary states can be found 
in Ref.~\cite{MMS},
where also the boundary state overlaps have been computed:
\begin{align}
_{\rm \tB}\!\langle k/4,0|q_c^{L_0-c/24}|k/4,1\rangle_{\rm \tB} &= \sum_{j\in\ZZ}\sum_{n\;{\rm odd}}\frac12\l[1+(-1)^{j+(n-1)/2}\r]\chi_{j,n,1},\no\\
_{\rm \tB}\!\langle k/4,0|q_c^{L_0-c/24}|k/4,-1\rangle_{\rm \tB} &= \sum_{j\in\ZZ}\sum_{n\;{\rm odd}}\frac12\l[1+(-1)^{j+(n-1)/2}\r]\chi_{j,n,-1}.\no
\end{align}
In the first overlap the multiplicity of $\chi_{(j,2j+1,1)}$ is one for all integer $j$. This shows that there are $k/2+1$ even Ramond ground
states. In the second overlap the the multiplicity of $\chi_{(j,-2j-1,-1)}$ is zero for all $j$. This shows that there are no odd
Ramond ground states. The Witten index is $(k+2)/2$ for both the brane and the anti-brane.

Now we want to match topological B-branes in the LG model with $W=x^{k+2}-y^2$ to Cardy B-branes in the minimal model ${\rm MM}_k$. 
We claim that for general $k$, the Cardy brane $|j,1\rangle_{\rm B}$ corresponds to the topological B-brane $E_{2j+1}$ in the LG model specified by
the following boundary tachyon profile:
$$d_0=\l(\begin{array}{cc}x^{2j+1}&y\\-y&-x^{k-2j+1}\end{array}\r), \quad
d_1=\l(\begin{array}{cc}x^{k-2j+1}&y\\-y&-x^{2j+1}\end{array}\r)$$
Up to reparametrization, there are $1+[k/2]$ different $E$'s. This is precisely the number of Cardy B-branes $|j,1\rangle_{\rm B}$. 
The open string spectrum  also matches up: the vector space ${\rm End}(E_{2j+1})$ has $2(2j+1)$ even basis elements and the same number of odd ones,
as we have seen in Section~\ref{sec:minimal}.

If $k$ is even, then the topological brane $E_{1+k/2}$ is a sum of two irreducible branes. The irreducible branes correspond to the following
factorization of $W$:
$$Q_{\pm}=\l(\begin{array}{cc}0&x^{1+k/2}\pm y\\x^{1+k/2}\mp y&0\end{array}\r).$$
We identify $Q_+$ with the Cardy brane $|k/4,1\rangle_{\rm \tB}$ and $Q_-$ with $|k/4,-1\rangle_{\rm \tB}$. One can easily check that their open 
string spectra match up as well.

\subsection{Cardy B-branes in ${\rm MM}_k/\ZZ_2$}

The B-type Ishibashi states in the orbifold theory are given by A-type Ishibashi states in ${\rm MM}'_k$ which are fixed by $\ZZ_{k+2}$. 
The Cardy states are given by
$$|j,s\rangle_{\rm B} = \sqrt{k+2}\sum_{2j'+s'\in2\ZZ}
\frac{S_{j,-2j-s,s}^{j',0,s'}}{\sqrt{S_{0,0,0}^{j',0,s'}}}\,|{\rm B};j',0,s'\drangle.$$
The overall normalization factor is fixed by requiring the ``vacuum'' representation 
to appear once in the partition function. 
Independent states come from 
$$2j=0,1,\ldots,\l[\frac{k}2\r], \qquad s=-1,0,1,2.$$
By adopting a convention for boundary condition for $G$, one can restrict the range for $s$ to $s\in\{-1,1\}$. 
Note that the boundary states contain RR pieces, and therefore the branes are orientable. The only effect of changing $s$ from
$1$ to $-1$ is to change the sign of the RR piece. There is one special case though: for $k$ even and $j=k/4$ the RR piece
vanishes, and the values $s=1$ and $s=-1$ give same brane. Thus the brane with $j=k/4$ is unorientable. This gives a total of 
$k+1$ branes for any $k$ (counting separately branes and anti-branes).

The even open string spectrum in the Ramond sector is computed by
\bea
Z &=& _{\rm B}\langle j,0| q_c^{L_0-c/24} |j,1\rangle_{\rm B} \no\\
	&=& \frac12\sum_{2j'+s'\in2\ZZ}\!\frac{S_j^{\;j'}S_j^{\;j'}S_{j'}^{\;\tj}}
{S_0^{\;j'}} \chi_{\tj,\tn,-1}(q_o)\no
\eea
The multiplicity for the Ramond ground state $(j,n,s)=(\ell,-2\ell-1,-1)$ is given by
\bea
	n_{jj}^{\ell} &=& \frac12\sum_{2j'+s'\in2\ZZ}\frac{S_j^{\;j'}S_j^{\;j'}S_{j'}^{\;\ell}}{S_0^{\;j'}}\no\\
	&=& \sum_{j'=0}^{k/2}\frac{S_j^{\;j'}S_j^{\;j'}S_{j'}^{\;\ell}}{S_0^{\;j'}}\no\\
	&=& N_{jj}^{\ell}\no
\eea
Since that $2j\le k/2$,  there is a single Ramond ground state for each $\ell$ in the range
$$\ell\in\{0,1,\ldots,2j\}$$
In other words, there are $2j+1$ even supersymmetric Ramond ground states. We note that the multiplicity of the vacuum representation is 
always one. This implies that all these B-branes are irreducible. 

Similarly the odd Ramond states can be read off from 
\begin{equation*}
_{\rm B}\langle j,0| q_c^{L_0-c/24} |j,-1\rangle_{\rm B} = 
\frac12\sum_{2j'+s'\in2\ZZ}\!\frac{S_j^{\;j'}S_j^{\;j'}S_{j'}^{\;\tj}}
{S_0^{\;j'}} \chi_{\tj,\tn,1}(q_o)\no
\end{equation*}
The multiplicity of the ground state $(j,n,s)=(\ell,2\ell+1,1)$ is also $N_{jj}^{\ell}$. This shows that the number of odd Ramond ground states
is also $2j+1$. In particular, the Witten index vanishes for all these branes.

Now we can match the Cardy branes $|j,\pm1\rangle_{\rm B}$ to topological B-branes in the LG model with $W=x^{k+2}$. This is
straightforward:
\bea
	|j,1\rangle_{\rm B} &\lr& E_{1+2j} = \l\{Q=\l(\
	\begin{array}{cc}0&x^{2j+1}\\x^{k-2j+1}&0\end{array}\r)\r\}\no\\
	|j,-1\rangle_{\rm B} &\lr& \bE_{1+2j} = \l\{Q=\l(\
	\begin{array}{cc}0&x^{k-2j+1}\\x^{2j+1}&0\end{array}\r)\r\}\no
\eea
Note that for even $k$ and $j=k/4$, $E_{k/2+1}$ is isomorphic to its own anti-brane $\bE_{k/2+1}$, which is compatible with the 
fact that $|k/4,1\rangle_{\rm B}=|k/4,-1\rangle_{\rm B}$. 
One can easily see that the open string spectrum computed in Section~\ref{sec:minimal} agrees with the CFT computation.

\section{Discussion}

In this paper we have analyzed the topological sector in $\N=2$ minimal models with D-branes. We have
constructed solutions to the matrix factorization equation, and computed disk correlators. In the case of $A$-type minimal models,
the branes we have constructed exhaust the list of irreducible topological branes, and any topological brane
is isomorphic to the direct sum of irreducibles~\cite{Orlov}. In the case of $D$-type minimal models we do not
have a proof that we have constructed all irreducible branes, but we suspect that we did. In the case of $E$-type minimal
models we have constructed some examples of irreducible topological branes, but we do not have any intuition about their total number
(we expect that it is finite). 

We have seen that adding a square to the superpotential $W$ has the same effect on branes as orbifolding by a $\ZZ_2$ symmetry.
In the CFT language, this orbifolding corresponds to changing the GSO projection from 0A to 0B, or vice versa. Although
we have demonstrated this only in the case $W=z^n$, we conjecture that the result holds in full generality.

In order to extend the computations in this paper to Gepner models, one has to generalize the formalism to LG orbifolds
(and orientifolds). The approach to brane-engineering used in this paper has an obvious ``equivariant'' version outlined
in the end of Ref.~\cite{ustwo}. What is lacking so far is a formula for topological correlators. In fact, even in the
closed string case there is no general formula which would compute topological correlators for LG orbifolds, and
existing computations are somewhat ad hoc. 

Another interesting direction to explore is the study of gravitational descendants in topological LG models with boundaries.
A complete understanding of gravitational descendants would enable one to compute the exact space-time superpotential
in Gepner models with D-branes. It would also enable one to understand how the category of D-branes is deformed
as one varies the superpotential $W$.

\section*{Appendix}

We present here disk correlators in $D$-type topological $\N=2$ minimal models for all the B-branes constructed in Section~\ref{subsec:D}. 
The algebra of topological closed string states has two generators $x$ and $y$ which satisfy the relations
$$
y^2 = (n+1)x^n, \quad xy=0.
$$
Clearly one can restrict to monomials $x^iy^j$, with $ij=0,\, 0\le j\le 1$ and $0\le i\le n$. 

The spectrum of B-branes and the corresponding algebras of boundary operators are analyzed in detail in Section \ref{subsec:D}. 
There we found two general types of branes, which were labeled $E$ and $E_{k,y}$. When $n\in2\ZZ$, there are two additional branes 
which were called $E_1$ and $E_2$. See Section \ref{subsec:D} for precise definitions of these objects and their boundary OPE algebras.

Let us start with the brane $E$. We take the most general operator insertions $x^iy^j\cdot a^\ell$ where $x^iy^j$ is a bulk operator as described 
above, and $a$ is the even generator of the boundary OPE which we called $a_1$ in Section~\ref{subsec:D}. By the results of Ref.~\cite{ustwo}, 
this disk correlator reduces to a multi-dimensional integral
$$\frac1{(2\pi i)^2}\oint_{L}\frac{-x^iy^{\ell+j+1}\,dx\wedge dy}{\part_1W\part_2W},$$
which can be readily computed by the method of residues. The result is
$$
\big\langle x^iy^j\cdot a^\ell\big\rangle_E \;=\; \Big \{ 
\begin{array}{ll} 1 & \qquad {\rm if}\;\; i=0, \;\ell+j=1\\0 & \qquad \;\; {\rm otherwise.}\end{array}
$$
In other words, there are only two nontrivial disk correlators: $\langle y\rangle_E = \langle a\rangle_E = 1.$ All other disk correlators 
vanish. 
The disk correlators for the anti-brane $\bar{E}$ are the same, except for a sign flip: 
$\langle y\rangle_{\bar E} = \langle a\rangle_{\bar E} = -1.$ Both $E$ and $\bar{E}$ carry RR charge.

For the branes labeled by $E_{k,y}$, it can be shown that unless $\xi\eta$ is inserted on the boundary, the supertrace vanishes. 
One can therefore restrict to insertions of the type $x^iy^j\cdot a^\ell\xi\eta$. Note that $a$ is not an independent generator but is given 
by $a=\eta^2$. 
See Section \ref{subsec:D} for various definitions. The disk correlator is given by the integral
$$\frac1{(2\pi i)^2}\oint_{L}\frac{-2(n+1)x^{n+\ell+i-k+1}y^{j}\,dx\wedge dy}{\part_1W\part_2W},$$
and the result is
$$
\big\langle x^iy^j\cdot a^\ell\xi\eta\big\rangle_{E_{k,y}} \;=\; \Big \{ 
\begin{array}{ll} 1 & \qquad {\rm if}\;\; j=0, \;\ell+i=k-1,\\0 & \qquad \quad\quad {\rm otherwise.}\end{array}
$$
Therefore the only nonvanishing disk correlators are those for $x^i\cdot a^\ell\xi\eta$ that satisfy the selection rule $\ell+i=k-1$, 
and they all have equal size. Clearly, these branes do not carry RR charge.

Lastly, we have two additional branes $E_1$ and $E_2$ if $n\in2\ZZ$. The general insertion is $x^iy^j\cdot a^\ell$, and the disk correlators 
are given by
\bea
	\big\langle x^iy^j\cdot a^\ell\big\rangle_{E_1} &=& 
	\frac1{(2\pi i)^2}\oint_{L}\frac{x^{\ell+i}y^{j}\big[(n+1)x^{n/2}-y\big]\,dx\wedge dy}{\part_1W\part_2W}\no\\
	&=& \Bigg \{ \begin{array}{rl} 1/2 & \qquad {\rm if}\;\; i=\ell=0, j=1,\\-1/2 & \qquad {\rm if}\;\; j=0,\; \ell+i=n/2,\\
 0 & \qquad \quad\quad {\rm otherwise.}\end{array}\no
\eea
In other words, the only nonvanishing correlators are 
$$\langle y\rangle_{E_1}=1/2, \quad \big\langle x^i\cdot a^{n/2-i}\big\rangle_{E_1}=-1/2, \quad i=0,\ldots,n/2.$$
The correlators for other brane $E_2$ are essentially the same except for a few sign changes:
$$
\langle y\rangle_{E_2} = 1/2, \quad \big\langle x^i\cdot a^{n/2-i}\big\rangle_{E_2}=1/2, \quad i=0,\ldots,n/2.$$
Recall that we showed earlier that $E_1$ and $E_2$ are neither isomorphic objects nor a brane-anti-brane pair. Now we see that this can also 
be inferred from their disk correlators. If they were isomorphic, their disk correlators would be exactly the same; if they were a 
brane-anti-brane pair, their correlators would be exactly opposite. Our computation suggests otherwise.

\section*{Acknowledgments}

We are grateful to Vladimir Baranovsky, Kentaro Hori, and Dmitri Orlov for useful conversations. 
This work was supported in part by the DOE grant DE-FG03-92-ER40701.

\end{document}